# Mobile Cloud Computing: A Comparison of Application Models

Dejan Kovachev, Yiwei Cao and Ralf Klamma
Information Systems & Database Technologies
RWTH Aachen University
Ahornstr. 55, 52056 Aachen Germany
{kovachev, cao, klamma}@dbis.rwth-aachen.de

*Abstract*—Cloud computing is an emerging concept combining many fields of computing. The foundation of cloud computing is the delivery of services, software and processing capacity over the Internet, reducing cost, increasing storage, automating systems, decoupling of service delivery from underlying technology, and providing flexibility and mobility of information. However, the actual realization of these benefits is far from being achieved for mobile applications and open many new research questions. In order to better understand how to facilitate the building of mobile cloud-based applications, we have surveyed existing work in mobile computing through the prism of cloud computing principles. We give a definition of mobile cloud coputing and provide an overview of the results from this review, in particular, models of mobile cloud applications. We also highlight research challenges in the area of mobile cloud computing. We conclude with recommendations for how this better understanding of mobile cloud computing can help building more powerful mobile applications.

*Index Terms*—cloud computing; mobile computing; remote execution; distributed systems; automatic offloading

## I. INTRODUCTION

Mobile devices allow users to run powerful applications that take advantage of the growing availability of built-in sensing and better data exchange capabilities of mobile devices. As a result, mobile applications seamlessly integrate with realtime data streams and Web 2.0 applications, such as mashups, open collaboration, social networking and mobile commerce [1], [2]. The mobile execution platform is being used for more and more tasks, e.g., for playing games; capturing, editing, annotating and uploading video; handling finances; managing personal health, micro payments, ticket purchase, interacting with ubiquitous computing infrastructures. Even mobile device hardware and mobile networks continue to evolve and to improve, mobile devices will always be resource-poor, less secure, with unstable connectivity, and with less energy since they are powered by battery. Resource poverty is major obstacle for many applications [3]. Therefore, computation on mobile devices will always involve a compromise.

Mobile devices can be seen as entry points and interface of cloud online services. Recently, has been discussed what cloud computing really means. Vaquero et al. [4] studied more than 20 definitions using the main characteristics associated with cloud computing. The cloud computing paradigm is often confused about its capabilities, described as general term that includes almost any kind of outsourcing of hosting and computing resources. According to NIST [5] cloud computing is a model for enabling convenient, on-demand network access to computing resources that can be rapidly provisioned and released with minimal management effort.

The combination of cloud computing, wireless communication infrastructure, portable computing devices, location-based services, mobile Web, etc., has laid the foundation for a novel computing model, called *mobile cloud computing*, which allows users an online access to unlimited computing power and storage space. Taking the cloud computing features in the mobile domain, we define:

*"Mobile cloud computing is a model for transparent elastic augmentation of mobile device capabilities via ubiquitous wireless access to cloud storage and computing resources, with context-aware dynamic adjusting of offloading in respect to change in operating conditions, while preserving available sensing and interactivity capabilities of mobile devices."*

For example, OnLive [6] executes video games in the cloud and delivers video stream to resource-poor clients without interrupting the game experience. Many other examples where the cloud can augment mobile devices can be envisioned, e.g. virus scan, mobile file system indexing, augmented reality applications.

In our survey we consider smart mobile devices that include devices continuously connected to the Internet, with handheld form and a large, high quality graphics display and significant but limited computing power.

To make this vision a reality beyond simple services, mobile cloud computing has many hurdles to overcome. Existing cloud computing tools tackle only specific problems such as parallelized processing on massive data volumes [7], flexible virtual machine (VM) management [8] or large data storage [9]. However, these tools provide little support for mobile clouds. The full potential of mobile cloud applications can only be unleashed, if computation and storage is offloaded into the cloud, but without hurting user interactivity, introducing latency or limiting application possibilities. The applications should benefit from the rich built-in sensors which open new doorways to more smart mobile applications. As the mobile environments change, the application has to shift computation between device and cloud without operation interruptions, considering many external and internal parameters. The mobile cloud computing model needs to address the mobile constraints in success to supporting "unlimited" computing capabilities for applications. Such model should be applicable to different scenarios. The research challenges include how to abstract the complex heterogeneous underlying technology, how to model all the different parameters that influence the performance and interactivity of the application, how to achieve optimal adaptation under different constraints, how to integrate computation and storage with the cloud while preserving privacy and security.

In this paper, we provide a systematic and comparative description of mobile application models that go along with the cloud computing ideas. The survey results show that current related projects cover only different subsets of the desired mobile cloud characteristics. Therefore, a new architecture design model for mobile applications that operates with the cloud needs to be adopted.

In rest of the paper, we go along the connection between mobile and cloud computing. We start by briefly checking the current status in mobile application models and their drawbacks (Section II). Next, we categorize, examine and compare different novel promising application models that fit to mobile cloud computing paradigm and compare them (Section III). In the next section we outline the research challenges ahead (Section IV). In the final section we conclude our paper.

## II. CURRENT STATUS IN MOBILE APPLICATIONS

Several researchers, [10]–[12], have identified the fundamental challenges in mobile computing. Mobile computing environments are characterized by severe resources constraints and frequent changes in operating conditions. Mobile devices inherently have and will continue to have limited resources as processing power, memory capacity, display size, and input forms. These have been the forming factors of exitsting mobile appliction approaches.

### A. Offline Applications

Most of the applications available for modern mobile devices fall into this category. They act as fat client that processes the presentation and business logic layer locally on mobile devices with data downloaded from backend systems. There is periodical synchronization between the client and backend system. A fat client is a networked application with most resources available locally, rather than distributed over a network as is the case with a thin client.

Offline applications, also often called native applications, offer:
- good integration with device functionality and access to its features
- performance optimized for specific hardware and multi-tasking
- always available capabilities, even without network connectivity

On the other hand, the native applications have many disadvantages:
- no portability to other platforms
- complex code
- increased time to market
- a requirement for developers to learn new programming languages

### B. Online Applications

An online application assumes that the connection between mobile devices and backend systems is available most of the time. Smartphones are popular due to the power and utility of their applications, but there are problems such as cross-platform issues. Here Web technologies can overcome them; applications based on Web technology are a powerful alternative to native applications.

Mobile have the potential to overcome some of the disadvantages of offline applications because they are:
- multi-platform
- directly accessible from anywhere
- knowledge of Web technologies is widespread among developers, greatly minimizing the learning curve required to start creating mobile applications

However, mobile Web applications have disadvantages:
- too much introduced latency for real-time responsiveness, (even 30 msec latency affects interactive performance [3])
- no access to device's features such as camera or motion detection
- difficulties in handling complex scenarios that require keeping communication session a over longer period of time

### C. Issues with Offline and Online Mobile Applications

Current applications are statically partitioned, i.e. most of the execution happens on the device or on backend systems. However, mobile clients could face wide variations and rapid changes in network conditions and local resource availability when accessing remote data and services. As a result, one partitioning model does not satisfy all application types and devices. In order to enable applications and systems to continue to operate in such dynamic environments, mobile cloud applications must react with dynamical adjusting of the computing functionality between the mobile device and cloud depending on circumstances. In other words, the computation of clients and cloud has to be adaptive in response to the changes in mobile environments [13].

## III. NOVEL APPLICATION MODELS FOR MOBILE CLOUD COMPUTING

Mobile cloud computing could be described as the availability of cloud computing services in a mobile ecosystem, i.e. world wide distributed storage system, exceed traditional mobile device capabilities, and offload processing, storage and security. To leverage the full potential of mobile cloud computing we need to consider the capabilities and constraints of existing architectures.

### A. Augmented Execution

Augmented execution refers to a technique used to overcome the limitations of smartphones in terms of computation, memory and battery. Chun and Maniatis [14] propose an architecture that addresses these challenges via seamlessly offloading execution from the phone to computational infrastructure (cloud) where *cloned replica* of the smartphone's software is running.

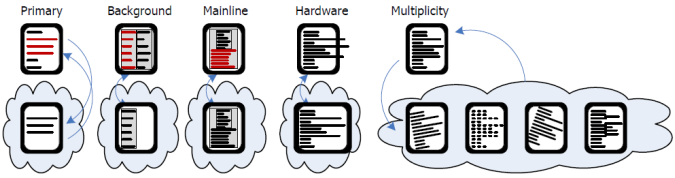

Fig. 1. CloneCloud categories for augmented execution (adapted from [14])

The mobile phone hosts its computation and memory demanding applications. However, some or all of the tasks are offloaded in the cloud where a cloned system image of the device is running. The results from the augmented execution are reintegrated upon completion. This approach for offloading intensive computations employs loosely synchronized virtualized or emulated replicas of the mobile device in the cloud. Thus, it provids illusions that the mobile user has a more powerful, feature-rich device than actually in reality, and that the application developer is programming such powerful device without having to manually partition the application or provision proxies. Instantiating device's replica in the cloud is determined based on the cost policies which try to optimize execution time, energy consumption, monetary cost and security.

Fig. 1 shows categorization of possible augmented execution for mobile phones: (1) primary functionality outsourcing - more like a client-server application, (2) background augmentation - good for independent separate process that can run in background like a virus scanning, (3) mainline - in-between primary and background augmentation, (4) hardware - the replica runs on more powerful emulated VM, and (5) multiplicity - helpful for parallel executions.

Similar approach of using virtual machine (VM) technologies executing the computation intensive software from mobile device is presented by Satyanarayanan et al. [3]. In this architecture, a mobile user exploits VMs to rapidly instantiate customized service software on a nearby cloudlet and uses the service over WLAN. A *cloudlet* is a trusted, resource-rich computer or a cluster of computers well connected to the Internet and available for use by nearby mobile devices. Rather relying on a distant cloud, the cloudlets eliminate the long latency introduced by wide-area networks for accessing the cloud resources. As a result, the responsiveness and interactivity on the device are increased by low-latency, one-hop, high-bandwidth wireless access to the cloudlet. The mobile client acts as thin client, with all significant computation occurring in a nearby cloudlet. This approach relies on technique called dynamic VM synthesis (cf. Fig. 2). A mobile device delivers small VMs overlay to the cloudlet infrastructure that already owns the base VM from which this overlay was derived. The infrastructure applies the overlay to the base to derive the VM which starts executing in the precise state in which it was suspended. However, Satyanarayanan et al. [3] report that the VM synthesis takes 60 to 90 seconds, which might not be acceptable for performing simple or ad hoc tasks. Garriss et al. [15] use a similar principle of running own VMs on public kiosks in order to establish a trustworthy and personalized computing environment. The user leverages a personal mobile device to gain degree of trust in a kiosk prior to using the kiosk. Using VMs enables the user to resume a complete personal computing environment that includes own choices of operating system, applications, settings, and data.

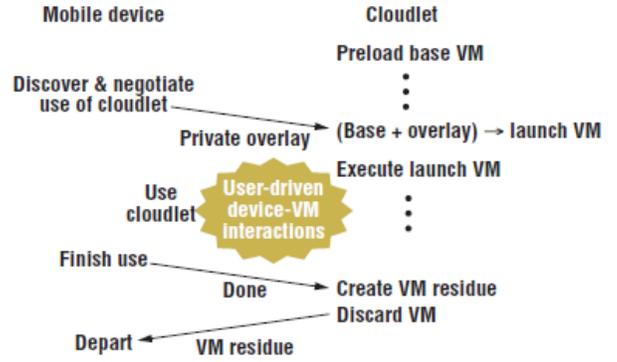

Fig. 2. Dynamic virtual machine synthesis timeline (adapted from [3])

### B. Elastic Partitioned/Modularized Applications

Running applications in heterogeneous changing environments like mobile clouds requires dynamic partitioning of applications and remote execution of some components. Applications can improve their performance by delegating part of the application to remote execution on a resource-rich cloud infrastructure.

Giurgiu et al. [16] develop an application middleware that can automatically distribute different layers of an application between the device and the server while optimizing several parameters such as latency, data transfer, cost, etc. In the core of this approach is a distributed module management which automatically and dynamically determines when and which application modules should be offloaded, in order to achieve the optimal performance or the minimal cost of the overall application. Giuriu et al. use the AlfredO [17] framework to carry out the distribution of the application modules between the mobile phone and the server. The AlfredO framework allows developers to decompose and distribute the presentation and logic layer of the application, while the data layer always stays on the server side. The minimal requirement is the UI of the application to run on the client side. Furthermore, Rellermeyer et al. [18] showed how such a modular application model enables elasticity. Elasticity in software can be observed as the ability to acquire and release resources on demand. Modules are units of encapsulation and units of deployment that compose the distributed application. The underlying runtime module management platform hides most of the complexity of distributed deployment, execution, and maintenance.

AlfredO is based on R-OSGi [19], a conceptual extension of the OSGi middlware model, that allows decomposition of

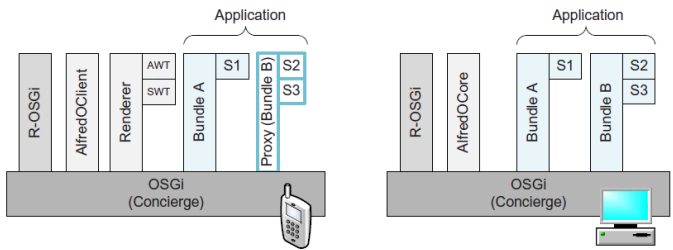

Fig. 3. AlfredO architecture (adapted from [16])

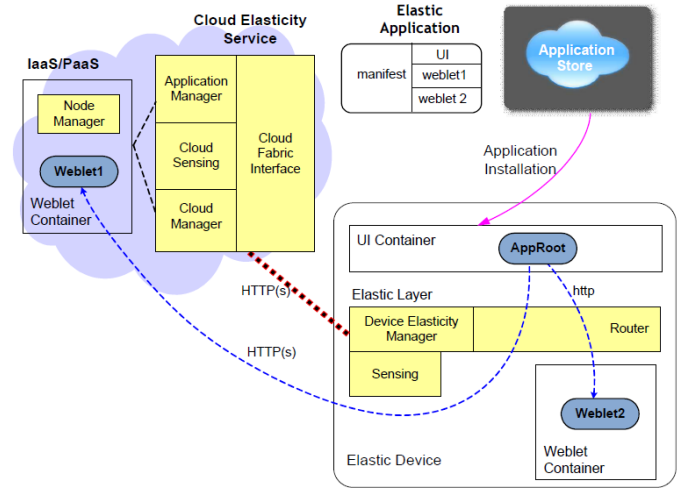

Fig. 4. Reference architecture for elastic applications (adapted from [20])

Java applications in *software modules*. A modified version of the original OSGi, namely R-OSGi, is used because the original OSGi allows only running services on the same Java virtual machine. Figure 3 shows the main concept. After the connection is established, the client requests an application. Then the optimal deployment for the application is computed. Based on that decision, an application description and a list of services to be fetched are sent to the client's Renderer. The Renderer generates corresponding UI according to the description. Furthermore, for the services that are decided to run on the client side the corresponding service bundles are fetched (on Fig. 3 service S1). Otherwise, for the services that are decided to run on the server, a local proxy on the client is created as an interface to this services (services S2 and S3).

Similarly, MAUI [21] is a system that enables fine-grained offload of mobile code to the cloud infrastructure. MAUI's goal is to maximize battery life of device with code offload. Developers annotate while programming which methods can be offloaded for remote execution. The profiling information for once offloaded methods is gathered, which is later used to better predict future invocations whether methods should be offloaded. The profiling information, network connectivity measurements, bandwidth and latency estimations are used as input parameters for an optimization problem which is periodically solved to give a decision which methods and when should be offloaded. Compared with [16], MAUI allows a fine grained offloading mechanism on the level of single methods, where in [16] the offloading happens on complete software modules. Even the experimental results from MAUI show that the separate method offloading can be contra-productive, i.e. several methods should be combined to achieve benefits.

Zhang et al. [20], [22] develop a reference framework for partitioning a single application into elastic components with dynamic configuration of execution. The components, called *weblets*, are platform independent and can be executed transparently on different computing infrastructures including mobile devices or IaaS (Infrastructure as a Service) cloud providers such as Amazon EC2 and S3 [8]. The application is split down to a UI component, weblets, and a manifest describing the application (cf. Fig. 4). Weblets are autonomous functional software entities that run on the device or cloud, performing computing, storing and network tasks. An elasticity manager component decides on migration, instantiation and migration of the weblets. This processes are transparent to the running application. The advantage of using such independent functional units - weblets - over AlfredO and R-OSGi is that weblets are not tied to one particular programming language or specification, allowing wider range of applications.

*C. Application Mobility*

The mobile cloud is accessed through heterogeneous devices. In order to provide seamless user experience same applications need to run on different devices. The application mobility plays a crucial role in enabling the next generation mobile applications. Application mobility is the act of moving application between hosts during their execution. Basically, application mobility is migrating running application states from one device to another to which the user has an immediate access [23], [24].

Application mobility is closely related to process migration. Process migration is an operating system capability that allows a running process to be paused, relocated to another machine, and continued there. It represents *seamless mobility* at the granularity of individual processes, and has been the research focus of many experimental projects [25]. However, application mobility involves more than process migration, e.g. migration tasks to different architectures or UI adaptation.

Satyanarayanan et al. [26] employ a mechanism called Internet Suspend/Resume (ISR), which allows one to logically suspend a machine at one Internet site, travel to some other sites and then seamlessly resume work there on another machine. ISR implementation is built on top of virtual machine technology and distributed file system. Each VM encapsulates distinct execution and user customization state. The distributed file system transports that state. However, one drawback is that migrating a complete virtual machine consumes more time and bandwidth than just selective application migration. Another drawback is that this works only on one platform type, otherwise the the latency is too high. In contrast to ISR, David et al. [27] propose an adaptive application mobility solution based on Java-based platform that supports mobile agents across heterogeneous hardware (JADE). In this approach, their design solution migrates individual applications and supports adaptation.

*D. Ad-hoc Mobile Clouds*

An ad-hoc computing cloud represents a group of mobile devices that serve as a cloud computing provider by exposing their computing resources to other mobile devices. This type of mobile cloud computing becomes more interesting in situations with no or weak connections to the Internet and large cloud providers. Offloading to nearby mobile devices save monetary cost, because data charging is avoided, especially favored in roaming situations. Moreover, it allows creating computing communities in which users can collaboratively execute shared tasks.

Huerta-Canepa and Lee [28] present guidelines for a framework to create virtual mobile cloud computing providers. This framework mimics a traditional cloud provider using nearby mobile devices. The proposed approach allows avoiding a connection to infrastructure-based cloud providers while bringing benefits of computation offloading. However, such an approach requires the support for spontaneous interaction networking with discovery and selection of mobile peers. Hadoop[1] ported on mobile device is used for distributing of processing tasks and storage. Communication is based on the Extensible Messaging and Presence Protocol (XMPP). The Hyrax project [29] employs a similar approach of using the Hadoop framework on mobile devices to share data and computation. Hadoop implements much of the core functionality needed for ad-hoc clouds, including global data access, distributed data processing, scalability, fault-tolerance, hardware interoperability and data-local computation. Since Hadoop is mainly designed for deployment on many servers, the major problem is how to enable the Hadoop framework to run on a mobile device.

Cao et al. [30] present a middleware that allows access from mobile devices to a bundle of multimedia services exposed from other mobile nodes. Mobile nodes can host web services that are accessed by other mobile nodes, thus exposing their computing capacities to the other mobile peers in an ad-hoc cloud. Particularly, related to ad-hoc clouds, much research in mobile ad-hoc and sensor networks has been done up to date.

*E. Comparison of Mobile Cloud Application Models*

A comparison of existing approaches for mobile cloud computing may point out the way to a better solution for mobile applications. The aforementioned application models fulfill in different scales the vision of mobile cloud computing. We have compared the models according to:

- *Middleware:* The enabling underlying technology used to achieve desired system properties.
- *Cost Model:* Are the different parameters of mobile clouds used to provide best performance?

[1] http://hadoop.apache.org

- *Programming Abstraction:* How powerful are the used programming tools to achieve quicker solid applications, while preserving the control over different mobile cloud parts?
- *Solution Generality:* Does the solution work for all applications or only for a few?
- *Implementation Complexity:* How difficult is it to develop mobile cloud applications?
- *Static & Dynamic Adaptation:* What is the separation of responsibilities between mobile clients and the cloud?
- *Network Load:* How large is the volume of data transferred? What is the introduced latency by offloading?
- *Scalability:* Can the application scale?

Table I shows how each of the approaches maps to the above attributes. The approaches from Cuervo et al. [21] and Zhang et al. [20] received top scores, because their model incorporates a cost model for deciding best execution configuration, the execution can also adapt dynamically. They provide a SDK that simplifies the development, and applications can scale both vertically and horizontally. The approach in [16] is similar, but lacks of dynamic adaptation of the computation between mobile devices and the cloud. Cloudlets [3] and ISR [26] allow high abstraction and personalization of the computing environment by using VMs, but lack from fine-grained execution adaptation. [28] and [29] approaches enable high horizontal scaling of the available ad-hoc mobile nodes, but with high communication overhead.

### IV. TOPICS FOR EXPLORATION

To enable the new mobile cloud application model, many challenges exist in different areas, including data replication, consistency, transaction management, cache management, optimal cost-effective execution in heterogeneous computing environments, elastic module lifecycle management and their communication and state synchronization. The middlware should provide an infrastructure for seamless and transparent execution of elastic applications and offer convenient development support.

*A. Programming Abstraction*

Development on mobile clouds should be simple and intuitive, however, at the same time the developer should be able to control behavior and location of his application. To take full advantage of modern mobile devices and available cloud computing resources, new programing abstraction tools hiding the complexity of underlying cloud technologies are needed. These tools need to raise the level of abstraction for application development, but enable also getting the performance on mobile client as described by Catanzaro et al. [32]. The developed software modules should be optimized for running on different mobile device hardware. Moreover, the programming tools need to support scalability and generate the cloud code, similar to MapReduce [33], a framework for doing batch processing jobs on thousands of machines. The programming tools should allow a holistic application development for the mobile client, middlware, and cloud, with dynamic shifting of the computation and the storage between them. For example, Zhang et al. [20] have implemented SDK, which is used to develop the basic interfaces of application modules and manage their lifecycle. Using the SDK, developers can build applications in high-level languages such as Java or C#. To support the rise of mobile clouds, not only new applications and services are of interest, but also the migration of existing applications and services to the cloud infrastructure. Coign [34] provides automatic program partitioning without source code modification.

TABLE I
COMPARISON OF EXISTING AND PROPOSED MOBILE CLOUD COMPUTING APPROACHES

| Application Model | Underlying Technologies | Cost Model | Programming Abstraction | Solution Generality | Implementation Complexity | Static Adaptation | Dynamic Adaptation | Network Load | Scalability |
|---|---|---|---|---|---|---|---|---|---|
| Offline | Vendor SDK | / | / | medium | low | high | low | high | low |
| Online | Web services, HTML5.0 | / | high | low | low | high | high | medium | high |
| Chun and Mantiatis [14] (CloneCloud) | DalvikVM (Android) | in [31] | / | low | high | high | low | low | high |
| Satyanarayanan et al. [3] (Cloudlets) | VirtualBox Dynamic VM synthesis | / | high | low | low | / | / | low | low (vertical) |
| Giurgiu et al. [16] (AlfredO and R-OSGi) | OSGi, Java | consumption graph | high | medium | low | high | low | low | medium (vertical) |
| Cuervo et al. [21] (MAUI) | .NET | linear optimization | high | high | low | high | high | low | high |
| Zhang et al. [20] (Weblets) | REST, C# | Naïve Bayes Classifier | high | high | low | high | high | / | high |
| Åhlund et al. [23] | P2P | / | / | low | low | medium | / | / | / |
| Satyanarayanan et al. [26] (ISR) | VM, Distributed File System | / | medium | low | low | low | low | high | medium (vertical) |
| Huerta-Canepa and Lee [28] | Hadoop, XMPP | / | high | low | low | medium | low | high | high (horizontal) |
| Cao et al. [30] (Mobile WS) | Web services | / | high | medium | low | medium | low | medium | medium (vertical) |
| Marinelli [29] (Hyrax) | Hadoop | / | high | low | low | medium | medium | high | high (horizontal) |

Alternatively, the complexity can be hidden in the middleware. For example, Wu et al. [35] propose an architecture that enables online, offline and mixed mode of operation for mobile applications with unified access to the business logic. The architecture is based on open standards which can be integrated with other platforms, and extended with other loosely coupled modules. The architecture framework enables easy building and adapting mobile applications that run in a selected mode depending on the scenario and user requirements. Different types of applications can be build or adapt without architectural changes. However, this approach is limited, in sense that dynamically changing the operation mode is not possible which requires rebuilding the application.

### B. Cost Model

In order to dynamically shift the computation between mobile device and cloud, applications needed to be split in loosely-coupled modules interacting with each other. The modules are dynamically instantiated on and shifted between mobile devices and cloud depending on the several metric parameters modeled in a cost model. These parameters can include the module execution time, resource consumption, battery level, monetary costs, security, or network bandwidth. A key aspect is user waiting time, i.e. that is the time a user waits from invoking some actions on the device's interface until a desired output or exception is returned to the user. User wait time is important for deciding whether to do the processing locally or remotely.

The cost model takes inputs from both device and cloud, and runs optimization algorithms to decide execution configuration of applications (cf. Fig. 5). Zhang et al. [20] use Naïve Bayesian Learning classifiers to find the optimal execution configuration from all possible configurations with given CPU, memory and network consumption, user preferences, and log data from the application. Guirgiu et al. [16] model the application behavior through a resource consumption graph. Every bundle or module composing the application has memory consumption, generated input and output traffic, and code size. Application's distribution between the server and phone is then optimized. The server is assumed to have infinite resources and the client has several resource constraints. The partitioning problem seeks to find an optimal cut in the graph satisfying an objective function and device's constraints. The objective function tries to minimize the interactions between the phone and the server, while taking into account the overhead of acquiring and installing the necessary bundles.

However, optimization involving many interrelated parameters in the cost model can be time or computation consuming, and even can override the cost savings. Therefore, approximate and fast optimization techniques involving prediction are needed. The model could predict costs of different partitioning configurations before running the application and deciding on the best one [31].

### C. Adaptation

Adaptation is key to mobility. Mobile cloud applications, running on relativly limited resources of mobile devices coping with unreliable networks and changing circumstances, must react on and dynamically reassign the responsibilities of mobile client and cloud, i.e. must be adaptive. In [10], the range of strategies for application and system adaptation is identified. The range is delimited by two extremes. At one extreme, adaptation is entirely the responsibility of individual applications. This approach, called laissez-fair adaptation, avoids the need for system support. The other extreme, called application-transparent adaptation, places the entire responsibility for adaptation on the system. A typical case of this approach is to use proxies to perform adaptation on behalf of applications. Between these two extremes lies a spectrum of possibilities that are referred to as application-aware adaptation. The application models from previous section lie somewhere between these boundaries.

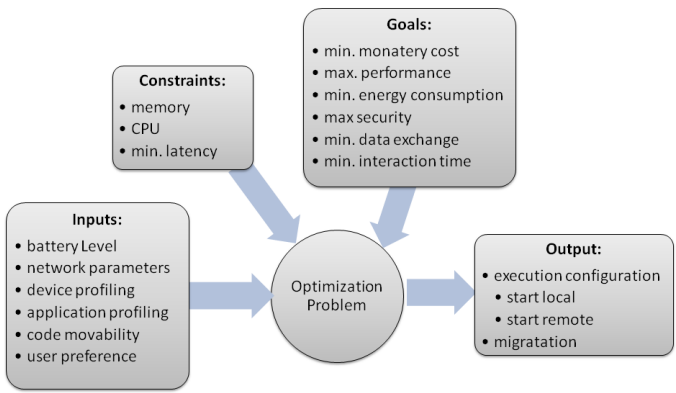

Fig. 5. Cost model of elastic mobile cloud applications (extended from [20])

### D. Cloud Integration

Cloud storage is the most obvious use of of cloud computing in mobile applications. Most devices have limited storage to hold applications, data, multimedia and operating system. The open questions that arise in this context are data transfer size optimization, and data persistence versus data availability [36]. Data transfer size optimization refers to how much data to move in a single transfer. Ideally the data transfer strategy should also have a degree of parameterization to handle stepping up and down the chunk size relative to network bandwidth, since bandwidth is highly variable in mobile applications. Data availability is important for completing tasks in a currently running process. Data persistence refers to storing data in the cloud until it is needed again in future. There is obviously a trade-off between them which requires taking into consideration of network connectivity, bandwidth, device capacity and latency. Caching can be used, but the use of cache on distributed databases requires additional efforts such as cache validation coherency.

Cloud processing offers a great option for offloading processing of large tasks requiring more time for calculation. However, many issues exist. First, cloud processing is more complex to implement than cloud storage because it involves both data management and synchronization. Second, to achieve truly computation augmentation of the limited mobile device, the performance of application execution needs to be monitored. For example, the supporting mobile cloud middleware can insert performance probes into the code and sense the running application modules. Third, moving software modules poses more issues. Some application modules can not be moved or it does not benefit from doing it. Moving stateless modules is more achievable than stateful modules.

Since data is shared on different systems, maintaining consistency becomes more and more important and difficult. Providing true transactional guarantees for software stacks that provides large scalability, anywhere, is an open gap for mobile cloud computing.

Currently, only little support is available to cross-platform execution and migration which mobile cloud computing structures will require. All of the analyzed approaches above are mostly tied to one specific middleware. Movement between cloud structures as a key issue has not been supported fully yet.

### E. Trust, Security and Privacy

A never ending issue will always be security in cloud computing related to multi-tenancy, concurrency, scale and distribution. First, direct concerns arise from aspects such as lacking control over data and code distribution in distributed infrastructures, potential data loss. Second, indirect issues arise from providing virtually unlimited computational resources to perhaps untrustworthy entities [37].

## V. CONCLUSIONS

In this paper, we have covered several representative mobile cloud approaches. Much other related work exist, but the purpose of this paper is to give an overview of the wide spectrum of mobile cloud computing possibilities. None of the existing approaches meets completely the requirements of mobile clouds. Native (offline) and Web (online) applications are the two extremes of mobile applications. The former type is using capabilities of mobile devices, but the integration with the cloud is poor. The latter type lacks from insufficient usage of mobile device sensors and available device computing resources while suffering from interactivity issues. Therefore, we believe that the full potential of mobile cloud applications lies in between these two extremes, while dynamically shifting the responsibilities between mobile device and cloud. Several researchers have shown how to achieve that by, e.g., replicating whole device software image or offloading parts of the application. The offloading can happen to some remote data center, nearby computer or cluster of computers, or even to nearby mobile devices. Moreover, due to the unstable mobile environments, many factors need to be incorporated in a cost model, and fast predictive optimizing algorithms decide upon the best application execution. To simplify the development a convenient, but effective, programming abstraction is required.

Mobile cloud computing will be a source of challenging research problems in information and communication technology for many years to come. Solving this problems will require interdisciplinary research from systems, networks, and HCI.